\documentclass[fleqn,usenatbib]{rasti}

\usepackage{newtxtext,newtxmath}

\usepackage[T1]{fontenc}

\DeclareRobustCommand{\VAN}[3]{#2}
\let\VANthebibliography\thebibliography
\def\thebibliography{\DeclareRobustCommand{\VAN}[3]{##3}\VANthebibliography}

\usepackage{graphicx}	
\usepackage{amsmath}	

\title[Seeing for PoET]{A seeing measurement device for the PoET solar telescope}

\author[B. Wehbe et al.]{
B. Wehbe,$^{1,2}$\thanks{E-mail: bwehbe@fc.ul.pt}
A. Silva,$^{3,4}$
M. Abreu,$^{1,2}$
A. Cabral,$^{1,2}$
N. C. Santos,$^{3,4}$
P. S\"{u}tterlin$^{5}$
\\
$^{1}$Instituto de Astrof\'isica e Ci\^encias do Espa\c{c}o, Universidade de Lisboa, Campo Grande, 1749-016, Lisboa, Portugal\\
$^{2}$Departamento de F\'isica, Faculdade de Ci\^encias, Universidade de Lisboa, Campo Grande, 1749-016, Lisboa, Portugal\\
$^{3}$Instituto de Astrof\'isica e Ci\^encias do Espa\c{c}o, Universidade do Porto, CAUP, Rua das Estrelas, 4150-762, Porto, Portugal\\
$^{4}$Departamento de F\'isica e Astronomia, Faculdade de Ci\^encias, Universidade do Porto, Rua do Campo Alegre, 4169-007, Porto, Portugal\\
$^{5}$Institute for Solar Physics , Dept. pf Astronomy, Stockholm University, AlbaNova University Center, SE 10691 Stockholm, Sweden
}

\date{Accepted XXX. Received YYY; in original form ZZZ}

\pubyear{\the\year{}}

\begin{document}
\label{firstpage}
\pagerange{\pageref{firstpage}--\pageref{lastpage}}
\maketitle

\begin{abstract}
Atmospheric seeing arises from stochastic fluctuations in the refractive index of the Earth’s atmosphere, producing random variations in the apparent direction of incoming light from astronomical sources. Scintillation refers to the associated intensity fluctuations induced by these refractive index inhomogeneities. A quantitative relationship between seeing and scintillation was established in 1993, enabling daytime seeing measurements by exploiting the Sun as an extended, bright source and using non-telescopic instrumentation. PoET, the Paranal solar ESPRESSO Telescope, will feed the Echelle SPectrograph for Rocky Exoplanets and Stable Spectroscopic Observations, ESPRESSO, at the European Southern Observatory (ESO) Very Large Telescope (VLT). By using the Sun as a proxy for solar-type stars, PoET will facilitate detailed investigations of the physical processes that drive stellar noise in ultra-high-precision radial-velocity measurements for exoplanet studies. The instrument is capable of targeting any region on the solar disk and acquiring spatially resolved spectra over areas ranging from 1 to 55 arcseconds. Accurate characterization of daytime atmospheric seeing is therefore essential for selecting the optimal observing aperture and ensuring the scientific performance of PoET. To support this requirement, we have developed and implemented a dedicated solar seeing monitor for daytime deployment at Paranal, Chile, where PoET will operate. In this work, we describe the instrument design and present the results from commissioning and initial on-sky validation.

\end{abstract}




\section{Introduction}

The performance of ground-based astronomical instruments is fundamentally constrained by the Earth's atmosphere. Although the twinkling of stars has been romantically celebrated in literature, astronomers regard it as a detrimental phenomenon that introduces image distortions and degrades observational quality. Two primary atmospheric effects influencing astronomical observations are dispersion and turbulence.

Atmospheric dispersion has been extensively studied in the context of its impact on ground-based instrumentation \citep{Wehbe2019, Wehbe2020a, Wehbe2020b}. It is now well established that the implementation of an Atmospheric Dispersion Corrector (ADC) is essential to mitigate this effect, e.g. ESPRESSO \citep{Pepe2021}, NIRPS \citep{Bouchy2025}.

In contrast, atmospheric turbulence represents a major limitation in optical astronomy, as it significantly reduces the achievable angular resolution of telescopes. Its impact on stellar observations can be categorized into two phenomena: seeing and scintillation. Seeing refers to random fluctuations in the direction of light entering the telescope, whereas scintillation describes stochastic variations in light intensity \citep{Schroeder2000}. In typical astronomical exposure times (few minutes), scintillation effects tend to average out.

A key parameter for quantifying seeing is the Fried parameter ($r_0$), which characterizes the spatial scale—often referred to as the coherence length — over which the atmospheric turbulence remains approximately constant \citep{Fried1965}. Within this region, wavefront aberrations of a light beam propagating through the atmosphere remain below approximately one radian. The advent of adaptive optics (AO) systems, capable of real-time correction of atmospheric wavefront distortions, has substantially enhanced the resolving power of ground-based telescopes (e.g., \cite{Rao2024, Davies2012, Hardy1998}). These advances have been instrumental in achieving higher spatial resolution and improving the fidelity of astronomical imaging.

While significant progress has been made in mitigating nighttime seeing, daytime seeing remains comparatively underexplored both in measuring and mitigation. Turbulent conditions during daylight are often exacerbated by solar heating, particularly in the near-ground layers, where the heating of the ground due to solar exposure lead to an increased turbulence. Consequently, continuous monitoring and characterization of daytime seeing are vital for solar astronomy. The Sun’s intense brightness and angular extent present unique observational challenges, with seeing effects directly influencing image quality and measurement accuracy. Previous studies have shown that daytime seeing can fluctuate substantially, underscoring the necessity of dedicated monitoring to understand and compensate for its impact on solar observations \citep{Griffiths2023}.

One of the most ambitious goals in modern astrophysics is the detection and characterization of Earth-like exoplanets—rocky worlds possessing physical conditions suitable for liquid water. Achieving this objective requires overcoming the dominant noise sources introduced by stellar astrophysical variability, including pulsations, granulation, and magnetic activity \citep{Klein2024, Santos2000}. Accurately modeling and mitigating these stellar signals are critical for the success of high-precision radial velocity (RV) programs and future missions targeting habitable-zone planets around nearby solar-type stars \citep{Hall2018, Thompson2016}.

The Sun serves as an ideal proxy for studying stellar activity and its impact on RV measurements. Accordingly, solar telescopes are being coupled to high-resolution spectrographs to investigate these effects by observing the disk integrated Sun (sun-as-a-star) \citep{Zhao2023, Haywood2016, Dumusque2015}. However, a detailed understanding of the underlying physical mechanisms—particularly those linked to granulation, oscillations, and magnetic activity—requires disk-resolved, high-precision solar spectroscopy.

The Paranal solar ESPRESSO Telescope (PoET, \citet{Santos2025}), currently under development at the Institute of Astrophysics and Space Sciences (IA) in Portugal, is designed to address this need. PoET will be linked to the ESPRESSO spectrograph \citep{Pepe2021} to deliver ultra-high-resolution (R > 200,000) spectra over a the optical wavelength window (380–780 nm). The telescope will be capable of targeting specific regions of the solar disk, acquiring spectra from areas spanning 1 to 55 arcseconds \textbf{(1, 2, 5, 10, 16, 29, 55)}. However, when observing small regions on the solar surface, image quality can be severely compromised by atmospheric seeing. Typical daytime seeing (measured in Paranal, Chile) corresponds to a Fried parameter ($r_0$) of 3–4 cm, equivalent to an angular resolution of approximately 2–3 arcseconds \citep{Griffiths2023, Ozisik2004}. Consequently, it is essential to equip PoET with a dedicated seeing measurement system to optimize daily observing strategies according to prevailing atmospheric conditions. In order to avoid contaminated spectra from regions outside of the defined aperture, the choice of the aperture will be informed by the measured seeing values.

In this context, we developed a seeing monitor based on the correlation between seeing and scintillation reported by \cite{Seykora1993}. The method has been proposed before by \cite{Beckers1993,Beckers1997} and used, in particular, for the ATST site selection and testing \citep{Beckers2003}. The instrument, known as the SHAdow Band Ranger (SHABAR) \citep{Sliepen2010} will enable continuous characterization of daytime seeing, thereby supporting the optimization of solar observations conducted with PoET.

\section{The instrument}

The SHABAR instrument provides a non-telescopic approach for quantifying atmospheric seeing conditions associated with extended sources, such as the Sun. The technique exploits the correlation established by \cite{Seykora1993} between solar irradiance fluctuations and optical seeing. By analyzing the covariance of intensity signals recorded from an array of photodiodes, SHABAR enables the retrieval of the vertical distribution of refractive index fluctuations above the instrument. This analysis yields the refractive index structure parameter profile, $C_n^2(h)$, as a function of height. An inversion algorithm is subsequently applied to transform the measured scintillation data into the corresponding $C_n^2(h)$ profile and to estimate the Fried parameter, $r_0$, which serves as a quantitative indicator of the prevailing seeing conditions. We point to \cite{Seykora1993} for details on the principle. \\
In the following subsections, we will detail the opto-mechanical design, the electronics design, and the software design of the instrument.

\subsection{Opto-mechanical design}
\label{subsec:optomec}
\begin{figure}
    \centering
    \includegraphics[width=1\linewidth]{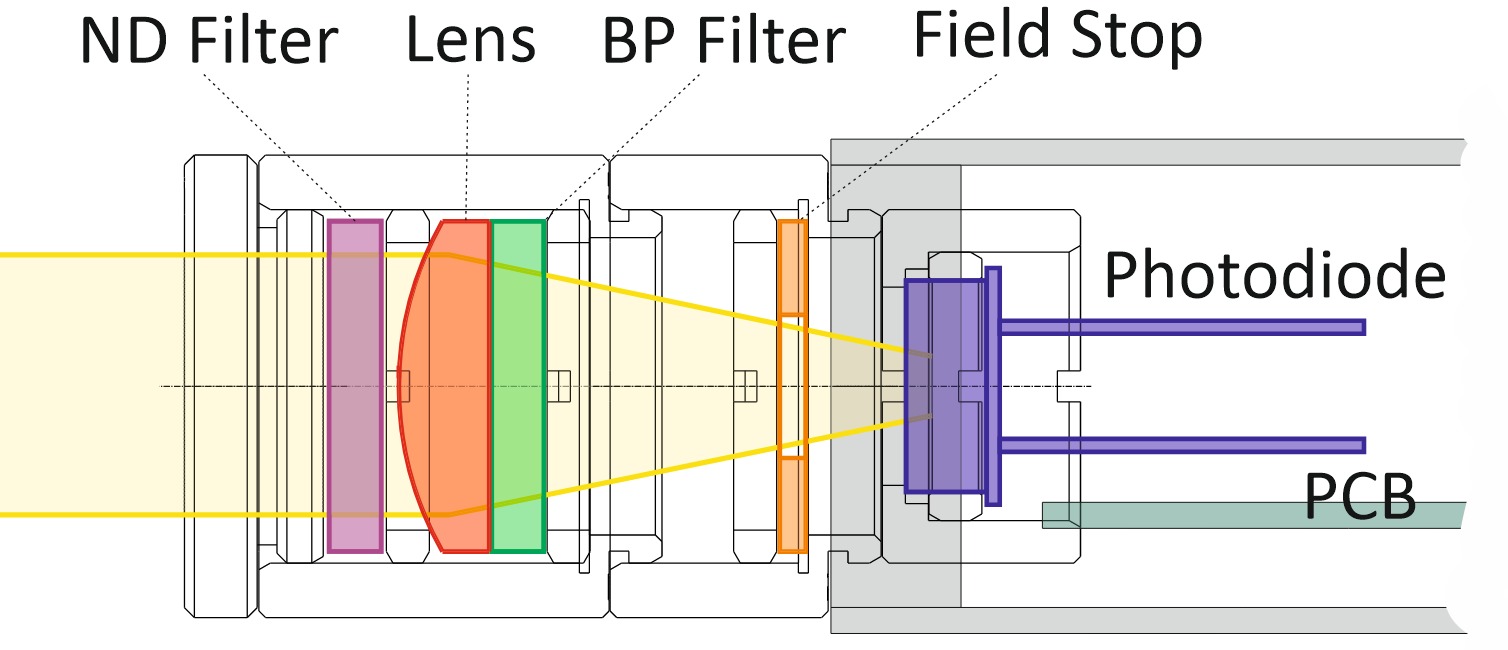}
    \caption{Optical layout of a scintillometer tube. The components are (from left to right): Neutral Density filter, lens, bandpass filter, field stop, photodiode. All components are off-the-shelf.}
    \label{fig:opticaltube}
\end{figure}

The opto-mechanical configuration of the SHABAR instrument closely follows the design described in \cite{Sliepen2010}. The system is comprised by six scintillometers mounted along a 50 cm bar, each positioned within apertures separated by distinct baselines (see Table \ref{tab:positions}). The correlation between the signal received by each pair of scintilometers separated by a different baseline $r$, is sensitive to turbulence at a specific range of altitudes. The choice of these separations is to maintain the same baseline as the instrument described in \cite{Sliepen2010}. Each scintillometer assembly includes a neutral density filter, a bandpass filter, a field stop, and an imaging lens (see Figure \ref{fig:opticaltube} and Table \ref{tab:scintillometer}), thereby enabling each unit to function as an independent optical channel. Solar scintillation is detected using photodiodes located at the focal planes of the imaging lenses. The bar assembly is mounted on a motorized telescope mount, which provides continuous solar tracking throughout the observing period.

\subsection{Electronics design}
The SHABAR optical sensors employ planar silicon PN photodiodes with integrated photopic filters. Each detection channel incorporates dedicated electronics for signal conditioning and amplification. To fully exploit the dynamic range of the 24-bit digitizer module, each sensor output is separated into alternating current (AC) and direct current (DC) components using complementary high-pass and low-pass filters with a crossover frequency of 1 Hz. The amplification gain of the AC channel can be adjusted to enhance sensitivity to solar flux fluctuations under conditions of minimal seeing. Each board has a jumper selector that allows changing the AC gain on 10X, 50X or 100X the DC gain, to allow exploring the full 24bit digitiser dynamics with maximum resolution.

\begin{figure}
    \centering
    \includegraphics[width=1\linewidth]{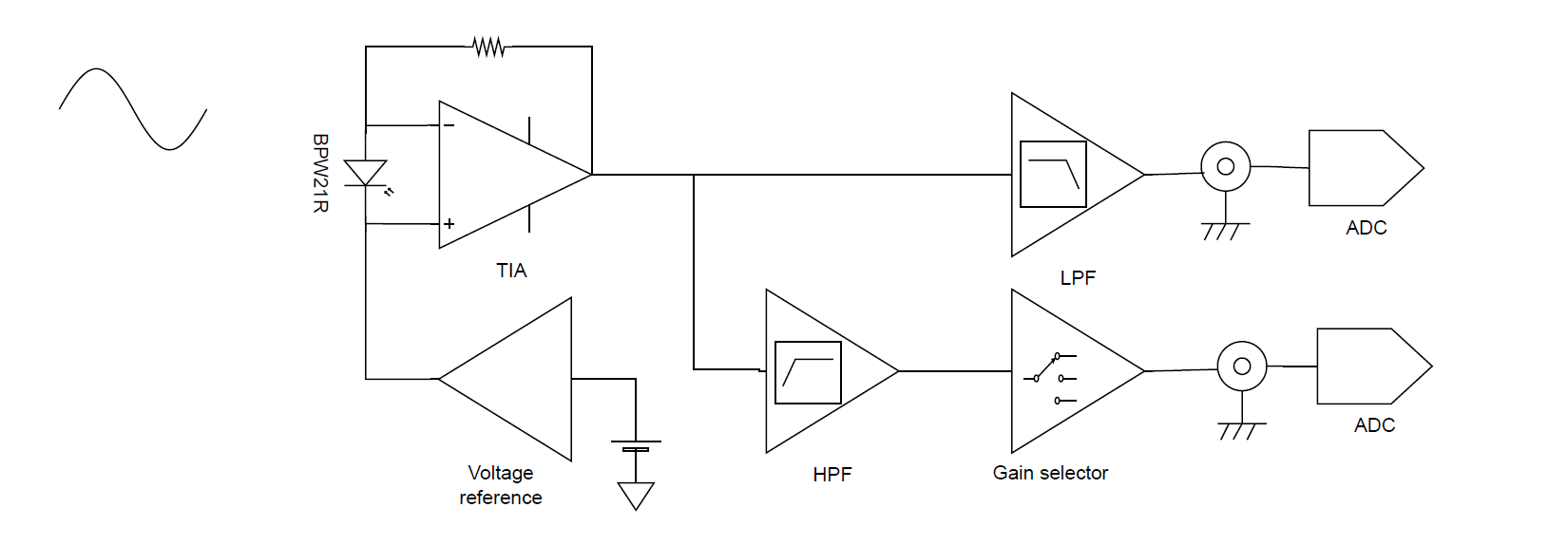}
    \caption{General schematic of the PCB showing its main functions. Note that a more detailed design can be accessed upon request from the authors.}
    \label{fig:pcb_schematics}
\end{figure}

The electronics associated with each channel are integrated onto compact printed circuit boards (PCBs), which are directly coupled to the corresponding optical sensors to minimize noise interference. Each PCB includes low-noise voltage regulators powered by an external supply. The first stage of amplification employs zero-drift transimpedance amplifiers to eliminate low-frequency drift effects during extended observation periods. The general schematic of the PCB is shown in Figure \ref{fig:pcb_schematics}.

The analog outputs from the six SHABAR scintillometers are connected to twelve single-ended analog input channels of the data acquisition board NI USB-6218 digitizer via high-quality, shielded coaxial cables, ensuring signal integrity and minimizing electromagnetic interference. \\

\subsection{Software design}

The estimation of the vertical profile of the optical turbulence strength, \( C_n^2(h) \), and associated atmospheric parameters is performed using time-series data of the AC and DC components acquired by the six SHABAR scintillometers over predefined integration intervals. The retrieval of the atmospheric turbulence profile is grounded in the theoretical framework of wave propagation through a random medium, as prescribed in \cite{Hill2003,Sliepen2010}. It is typically assumed that the atmosphere along the line of sight consists of multiple thin turbulent layers, and that the refractive index fluctuations follow a Kolmogorov spectrum, where temperature and humidity perturbations scale as the magnitude of the spatial wavenumber \( k^{-5/3} \).  

\begin{figure}
    \centering
    \includegraphics[width=1\linewidth]{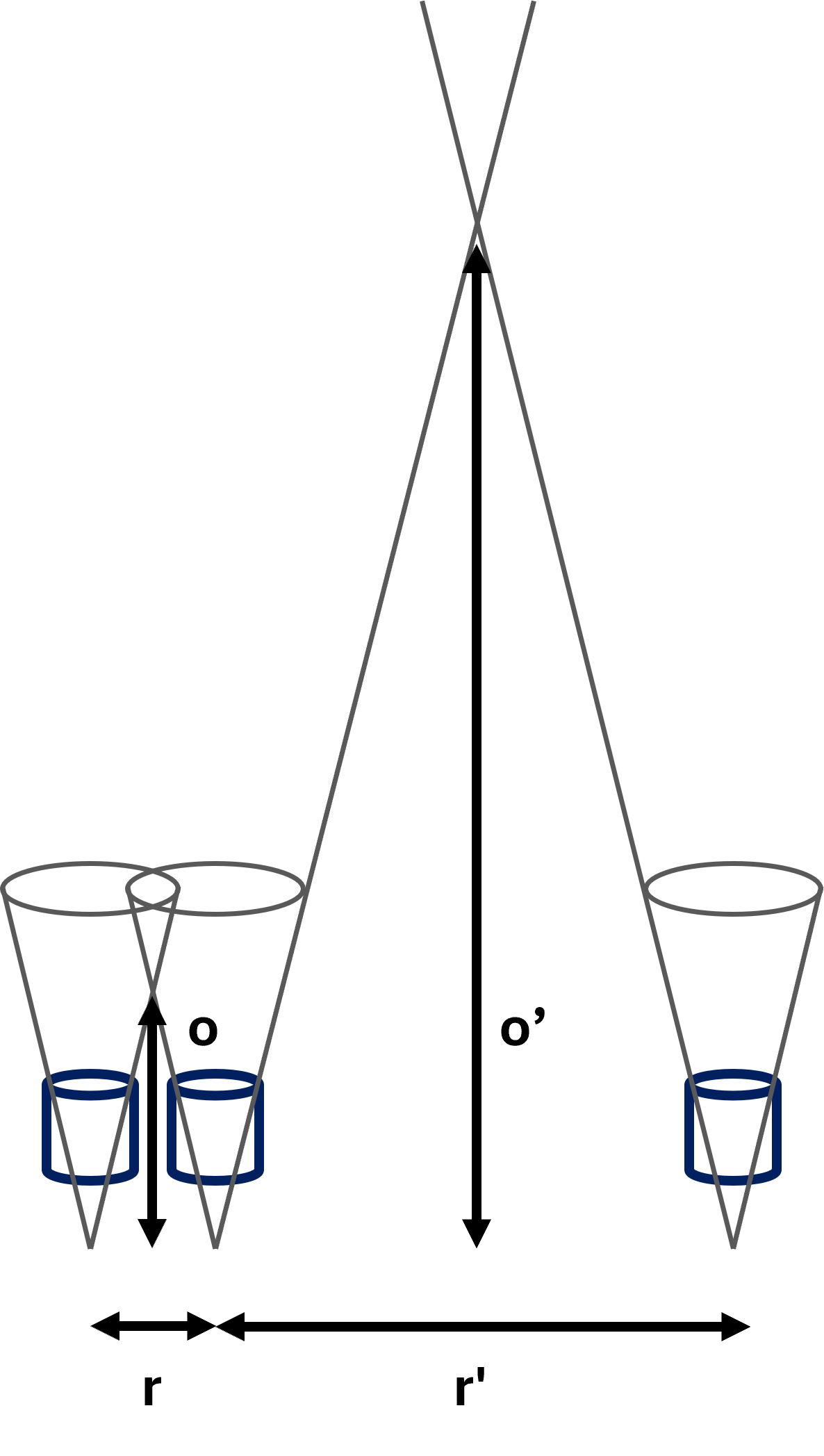}
    \caption{The geometry of the beams defined by the source and detectors intersecting the turbulence: $o$ and $o'$ are the height at which the cones from two detectors intersect. The separation between the detectors is denoted by $r$ and $r'$.}
    \label{fig:beams}
\end{figure}

The operational principle of the SHABAR relies on reconstructing the vertical distribution of turbulence strength along the line of sight by exploiting the overlapping fields of view of the individual detectors. The varying baseline separations between scintillometers result in intersection regions corresponding to different effective heights in the atmosphere (see Figure~\ref{fig:beams}). Based on the Kolmogorov turbulence model, a mathematical relationship can be established to map the \( C_n^2(h) \) profile to the covariance of intensity fluctuations measured at various separations, thereby enabling a model-based inversion of the observed correlations to recover the atmospheric turbulence profile.  

A detailed derivation of the covariance model and its dependence on altitude is provided in~\cite{Hill2003}; the principal computational steps are summarized as follows:  

\begin{enumerate}
    \item \textbf{Signal normalization:} Normalize the AC and DC components of each channel by their respective electronic gains.  
    \item \textbf{Correlation computation:} For each detector pair, select an appropriate averaging interval and compute the pairwise correlations. In our case, under nominal conditions, correlations between the AC signals of all detector pairs are evaluated every 2~seconds, yielding a total of 15 distinct covariance measurements.  
    \item \textbf{Kernel construction:} Construct the sensitivity kernels and define the altitude grid for the \( C_n^2(h) \) model. The grid is more finely sampled at lower altitudes, where the turbulence exhibits higher spatial variability.  
    \item \textbf{Model fitting:} Fit the parameters of an atmospheric turbulence model to the measured correlations. We use the same assumption as described in \cite{Sliepen2010}:
    \begin{itemize}
        \item A uniform-layer model assuming constant \( C_n^2(h) \) values within defined altitude intervals.  
    \end{itemize}
    \item \textbf{Seeing estimation:} Integrate the best-fit turbulence model to derive the Fried parameter, \( r_0 \), and compute the corresponding seeing value (at $\lambda = 520 \:nm$):  \\
    \begin{equation*}
        seeing = 0.98 \frac{\lambda}{r_0}
    \end{equation*}
    \item \textbf{Temporal monitoring:} Repeat the full inversion process continuously throughout the day, particularly during concurrent observations with the PoET telescope.  
\end{enumerate}
The zenithal angle of observation is taken into account in the computation of the Fried parameter $r_0$. A more detailed description of the software (algorithm, performance, limitations) will be presented in a dedicated paper (in preparation).

\section{Tests and Commissioning}
In this section, we will describe the tests done in the laboratory to characterize the different scintillometer units. A first set of characterization tests (mainly to quantify the response of the different scintillomer units) can be found in \cite{Wehbe2024}. We will focus on the gain characterization tests done in the lab as well as the commissioning tests done on-sky. 

\subsection{Laboratory tests}
\label{subsec:labtests}
As mentioned in \cite{Wehbe2024}, it is important to quantify the AC and DC gains of each scintillometer unit in order to understand if any normalization is necessary. In theory, all the printed circuit boards (PCB) are identical with a nominal $DC_{gain} = 15.000$ and a $AC_{gain} = 10, 50, 100 \times DC_{gain}$ based on the position of the jumper. \\
To characterize experimentally the gain of each scintillometer channel we had to implement a different setup to gauge the effective DC and AC gain. \\
The DC test setup used a precision source meter to inject a known current to the transimpedance amplifier (TIA) circuit. The total gain (V/A) was estimated from voltage level measured after the low pass filter amplifier. \\


The AC test setup used a wavefront generator producing a sinusoidal signal of 100 Hz, injected to the TIA through a decoupling capacitor. The exact AC gain was obtained by probing the TIA input current and the output voltage level for the different gain selector positions.


The results shown in Table \ref{tab:gains} validate the stability and uniformity of the scintillometrs channels.

\subsection{On-sky tests}
\begin{figure}
    \centering
    \includegraphics[width=1\linewidth]{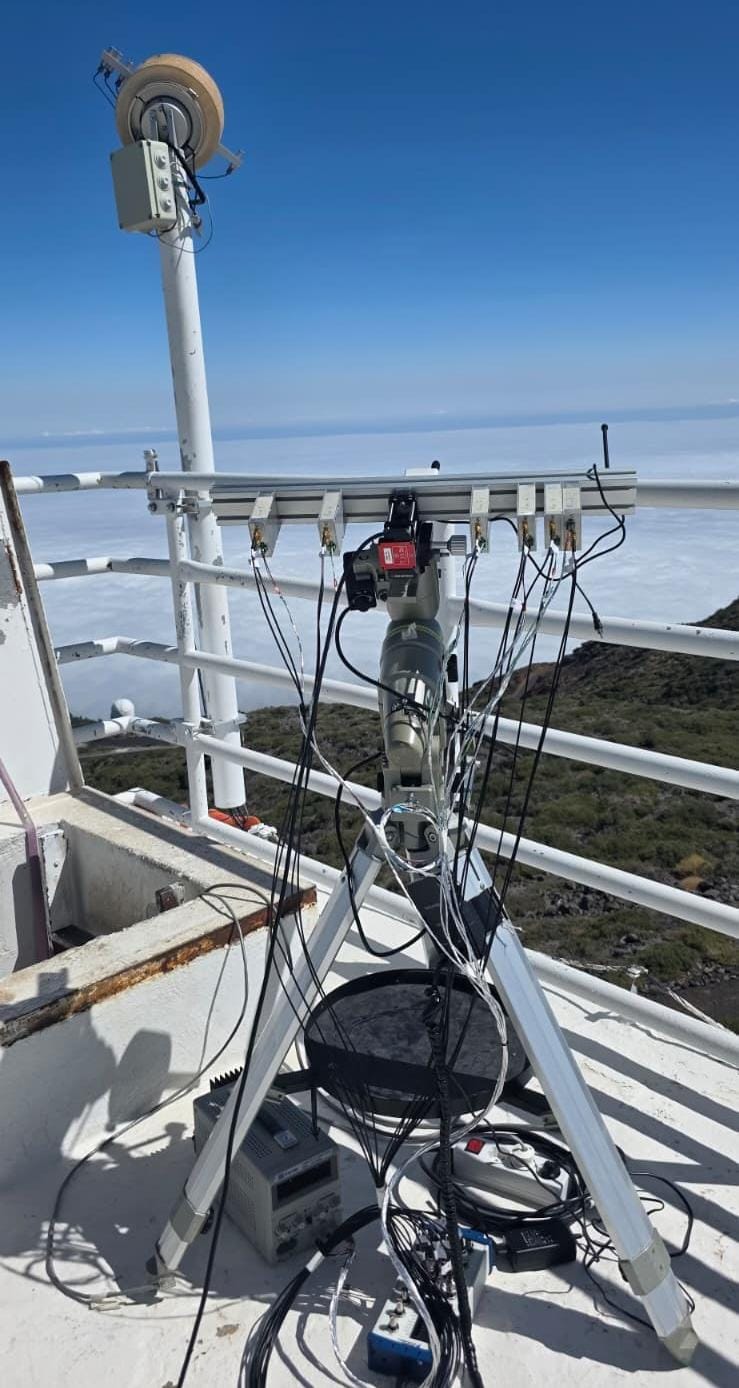}
    \caption{$SHABAR_{P}$ installed on a mount shown in the foreground of the picture. $SHABAR_{S}$ installed on a post in the background of the picture.}
    \label{fig:shabar-sst-poet}
\end{figure}

As mentioned in subsection \ref{subsec:optomec}, the design of our instrument (hereafter $SHABAR_{P}$) follows closely the one described in \cite{Sliepen2010} (hereafter $SHABAR_{S}$). The $SHABAR_{S}$ instrument is currently installed at the Swedish Solar Telescope (SST) at the Roque de los Muchachos Observatory, La Palma in the Canary Islands \citep{Scharmer2003} and has been running continuously, providing $r_0$ parameters and seeing measurements. In order to test the performance of our instrument, a commissioning campaign was held from July 26 to 31, 2025. Our instrument was assembled and installed at the top of the solar telescope tower (16 m above ground layer) where the $SHABAR_{S}$ is installed (see Figure \ref{fig:shabar-sst-poet}). It is important to state here that during that period, the SST was in a commissioning period where the telescope was not observing which allowed us easy access to the top of the tower. However, as we will show later, the fact that the telescope was being moved at random times during the day (for testing purposes) might have affected the results from one of the days. \\

\begin{figure}
    \centering
    \includegraphics[width=1\linewidth]{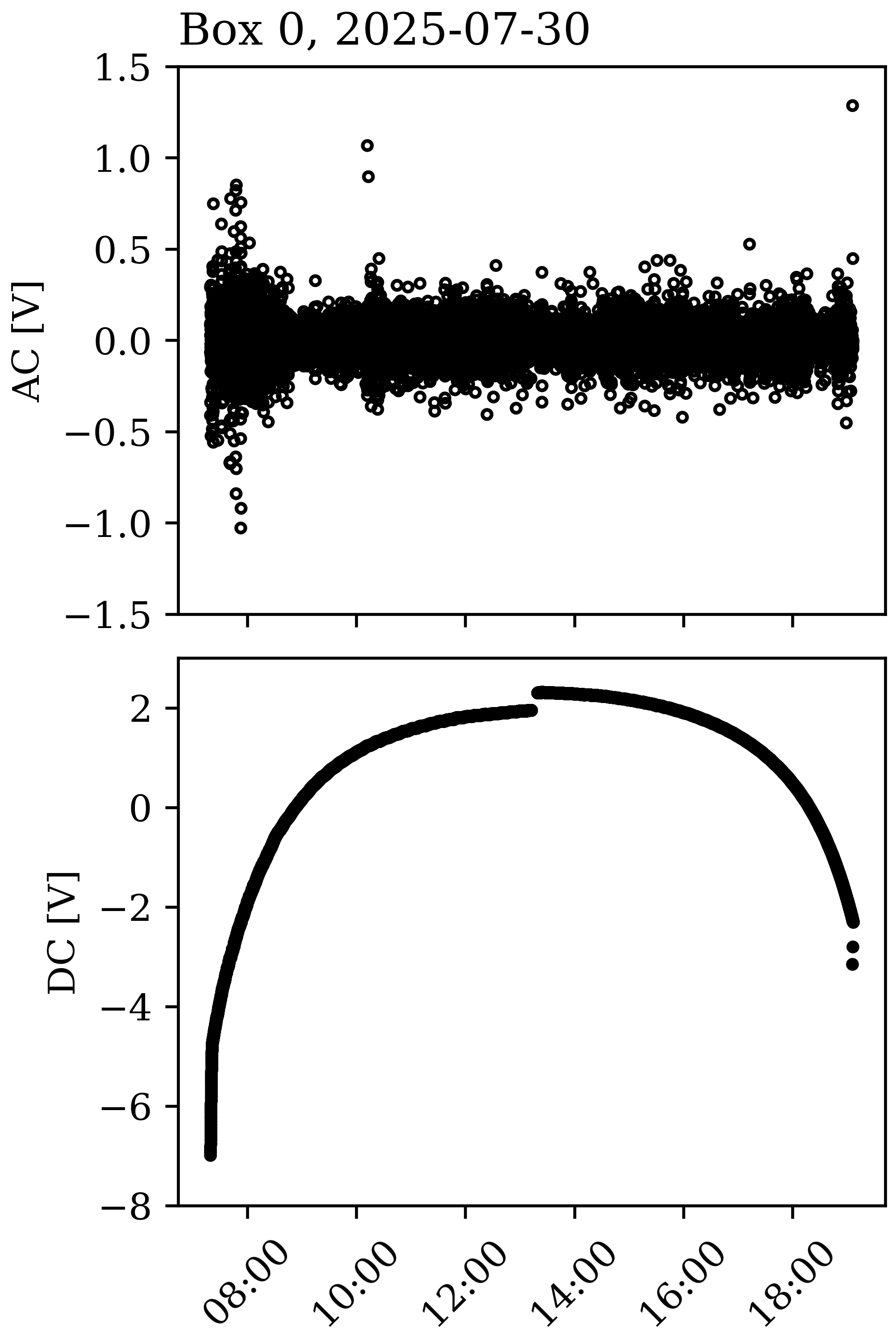}
    \caption{AC (top) and DC (bottom) raw signals over Box 0 for one day of observation (30$^{th}$ of July, 2025).}
    \label{fig:ACDC_box0_oneday}
\end{figure}

\begin{figure*}
    \centering
    \includegraphics[width=17cm]{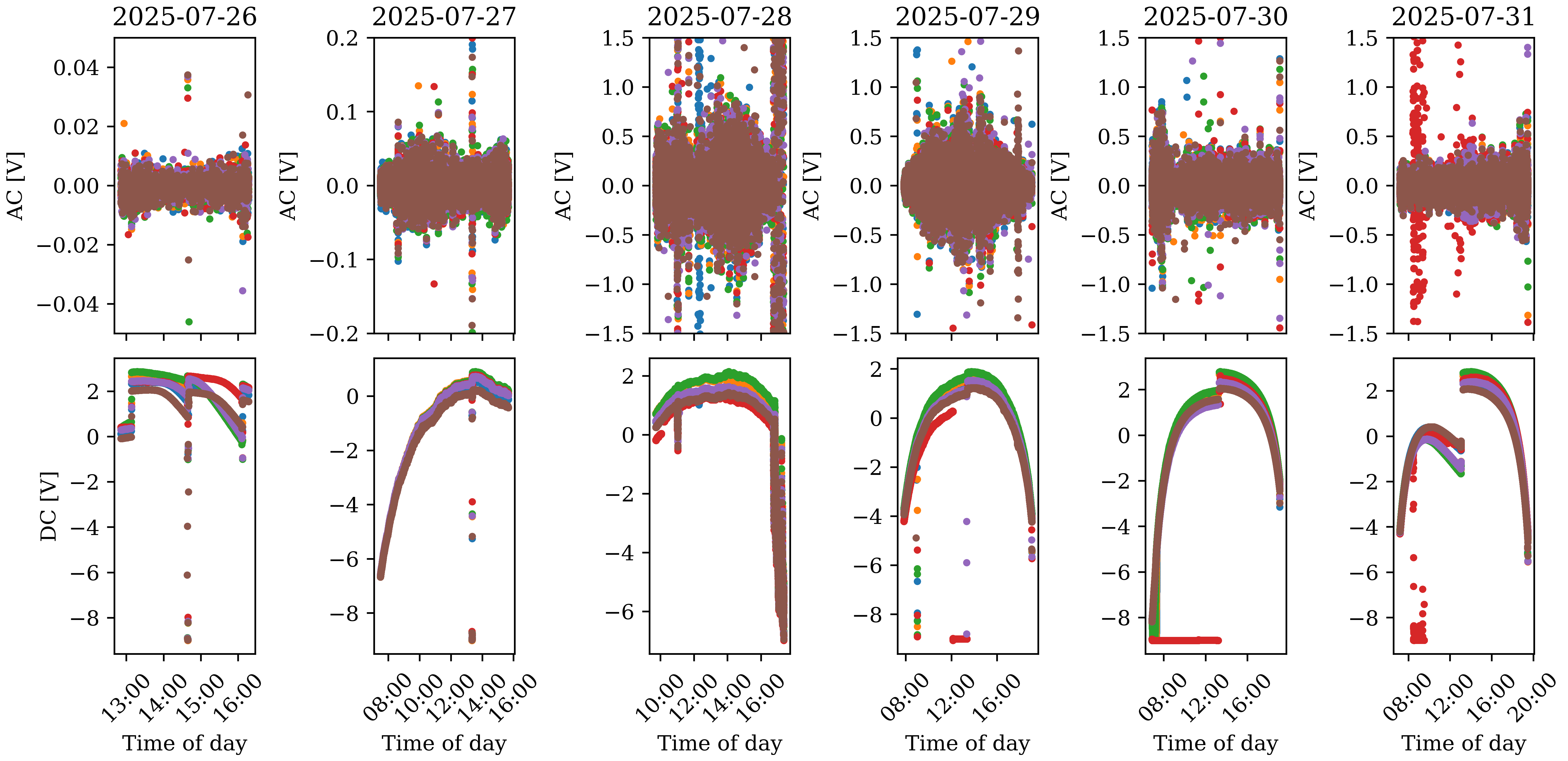}
    \caption{AC (top) and DC (bottom) raw signals over the full commissioning period. The difference in the amplitude of the AC signal is due to different jumper position across the days.}
    \label{fig:ACDC}
\end{figure*}

In order to keep the scintillometers pointing at the Sun all the time, the $SHABAR_{P}$ was installed on a motorized mount (Vixen GP DX) and controlled via an open source astronomical software N.I.N.A\footnote{\url{https://nighttime-imaging.eu/}}. A polar alignment was performed the night before the solar data acquisition to ensure that we had a proper solar tracking throughout the day of the $SHABAR_{P}$.
The output of each scintillometer is a DC and AC signal that are saved into a 2 seconds file at a sampling rate of 1000 Hz. The DC signal represents the intensity of the light and the AC signal represents the fluctuation (scintillation) of the light across the DC value. The raw data of one of the scintillometers (Box 0) across one day of observation (30$^{th}$ of July, 2025) is shown in Figure \ref{fig:ACDC_box0_oneday}. The behavior of the remaining of the boxes across the full commissioning period is similar. The raw DC and AC signals across the commissioning days are shown in Figure \ref{fig:ACDC}. On the first day ($1^{st}$ column of Figure \ref{fig:ACDC}) data was only collected after mid-day, hence the different time scale with respect to the rest of the days. The behavior of the DC signal is as expected: low values during the morning, increasing to reach maximum at meridian, and decreasing again in the afternoon as the sun is setting. Note that we had a malfunction in one of the boxes (Box 3) that could be seen in the DC plots associated with a value of -9 V. For consistency sake, we fully rejected the data from box 3 for the seeing calculation that will be presented later on. The jumps in the DC signal, more clear on the last two days, are due to the meridian flip where we had to stop, perform the flip of the mount manually, and resume data collection. The amplitude difference in the AC signal seen in Figure \ref{fig:ACDC}, is due to the different jumper selector position between the days. \\
The raw signals are then fed into the pipeline to compute the $r_0$ Fried parameter and hence a seeing value. The results are shown in Section \ref{sec:results}.

\section{Results}
\label{sec:results}
In this section, the results are divided into two subsections: our $SHABAR_P$ instrument results over the full commissioning period where we show the potential of the instrument; and the validation against the SST $SHABAR_S$ where we validate our pipeline and compare the performance of both instruments.

\subsection{$SHABAR_P$ results}
\begin{figure}
    \centering
    \includegraphics[width=1\linewidth]{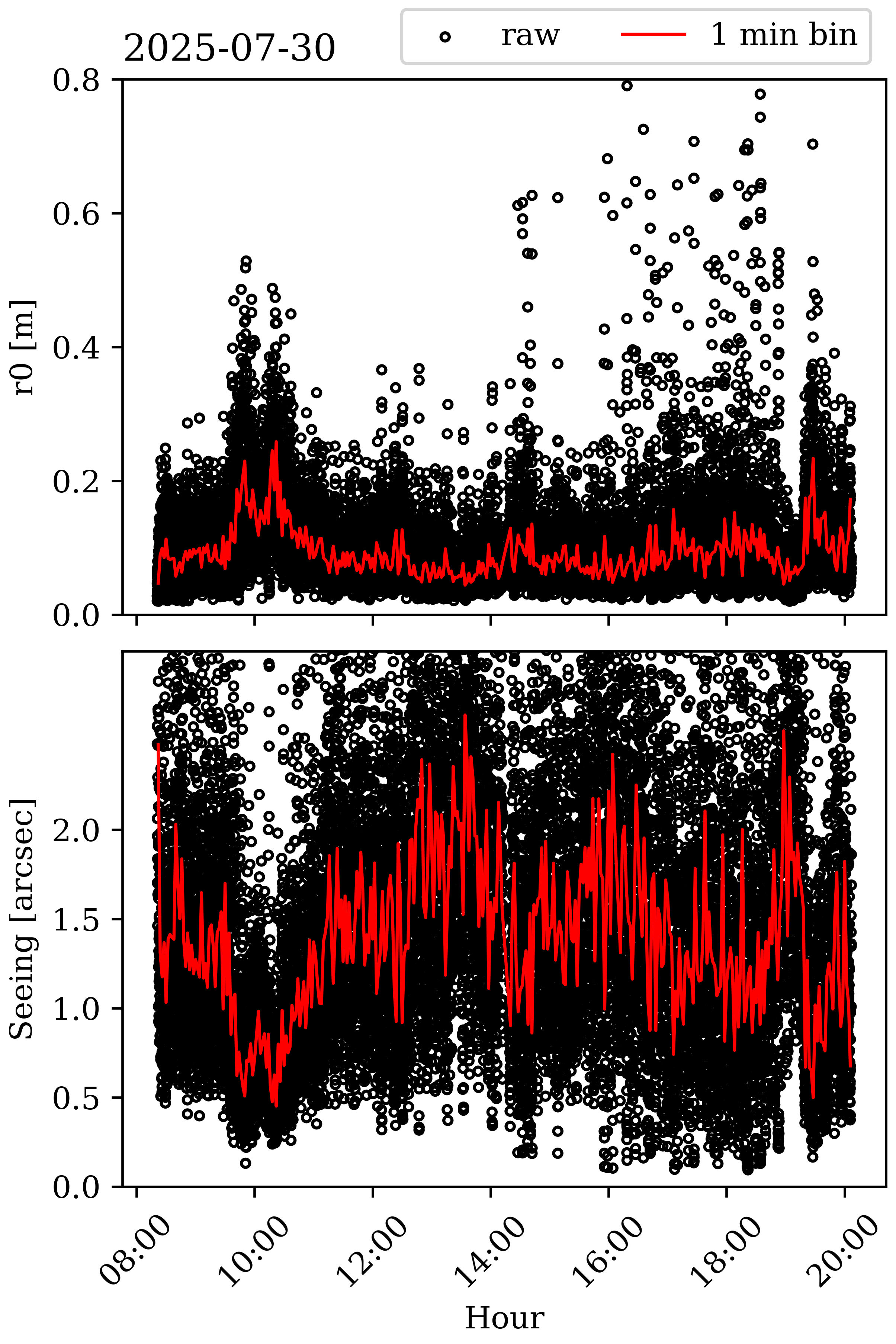}
    \caption{$r_0$ (top) and seeing (bottom) time series obtained over one day \textbf{(30$^{th}$ of July, 2025)}. The binned data over a 1-minute period is also shown.}
    \label{fig:unbinned}
\end{figure}

Figure \ref{fig:unbinned} shows the results of the measured $r_0$ parameter and seeing over the full day of the $30^{th}$ of July. For a better representation of the results, a 1-minute moving average has been applied to the data to get a smoother and more realistic values for $r_0$ and seeing.

\begin{figure}
    \centering
    \includegraphics[width=1\linewidth]{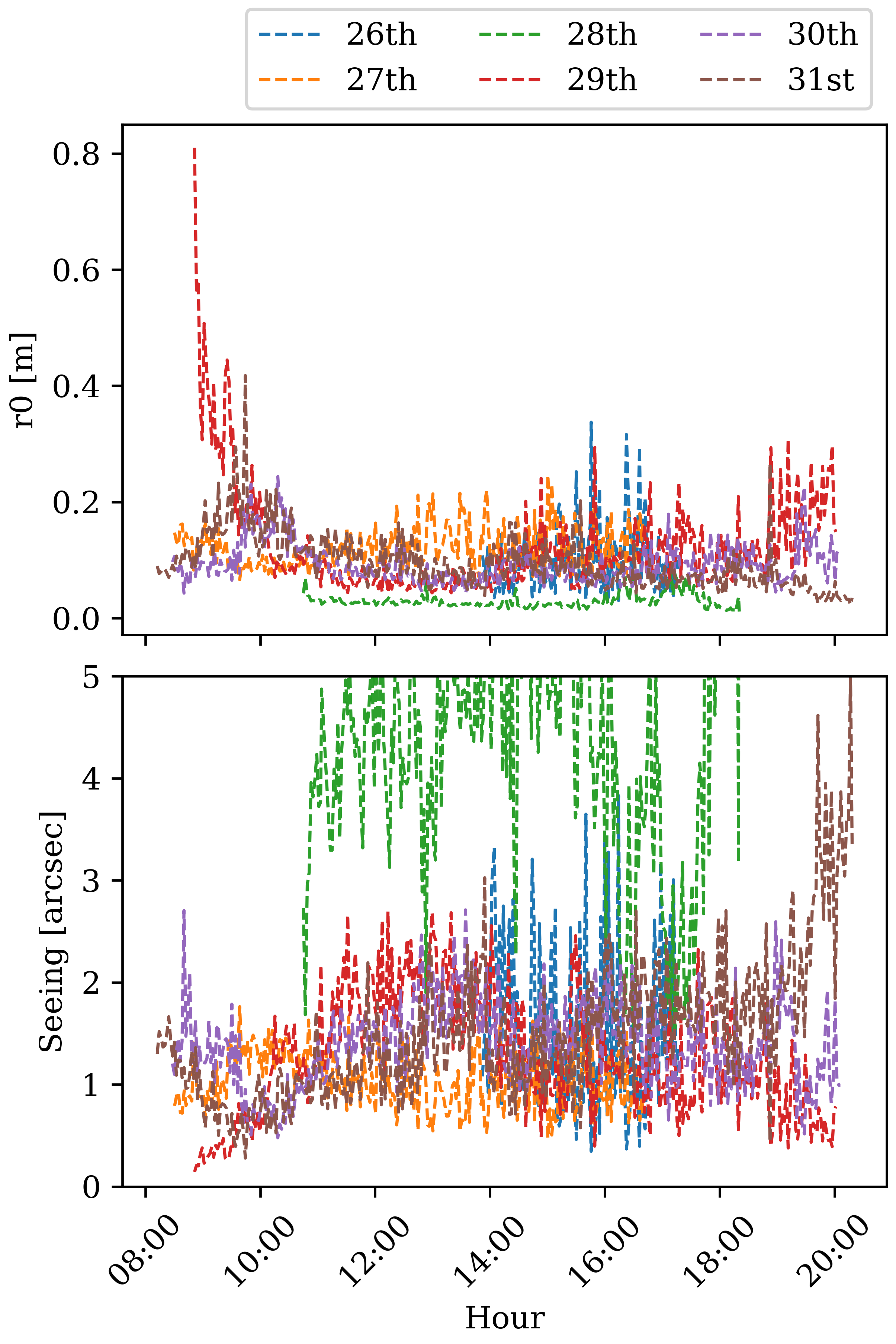}
    \caption{$r_0$ (top) and seeing (bottom) 1-minute binned time series obtained over the full commissioning period. The observation was done between the 26$^{th}$ and the 31$^{st}$ of July, 2025.}
    \label{fig:daily_fold}
\end{figure}

The results for all the commissioning days are shown in Figure \ref{fig:daily_fold}. With the exception of the $28^{th}$ (green curve at the beginning of the afternoon), the results from Figure \ref{fig:daily_fold} show a seeing average of 1.2 arcseconds over the day (with a minimum and maximum seeing of 0.7 and 2.2 arcseconds, respectively).  The noticeable increase in seeing values on these the $28^{th}$ can be attributed to the fact that the sky was very cloudy in the afternoon which was translated into higher seeing values. \\
The general tendency of the results show lower seeing values at the beginning of the day and at the end of it, with relatively higher values in between. This effect is expected as the higher the sun is, the more local thermal turbulence is induced that will affect the seeing values.

\subsection{Validation with SST data}
\label{subsc:validation}

\begin{figure}
    \centering
    \includegraphics[width=1\linewidth]{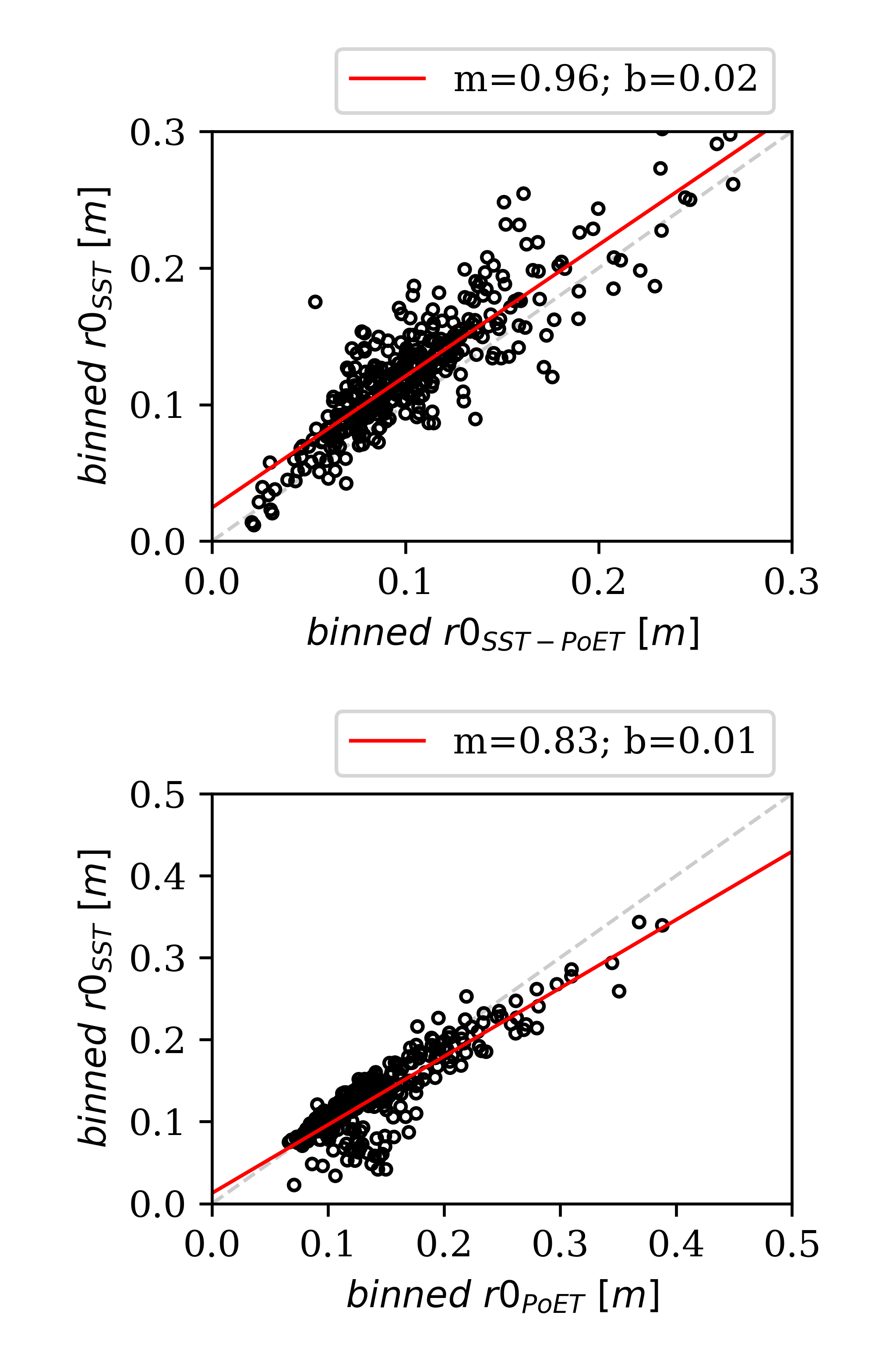}
    \caption{top pannel: scatter plot of the $SHABAR_S$ data reduced by the PoET pipeline vs $SHABAR_S$ reduced by the SST; bottom pannel: scatter plot of the $SHABAR_P$ data vs the $SHABAR_S$ reduced by the SST. We also show the regression line with the corresponding slope of each plot.}
    \label{fig:regression}
\end{figure}

The first test performed was to validate our pipeline against the SST data. We ran the covariances obtained from the $SHABAR_S$ into our pipeline. The choice of using the covariances and not the raw DC and AC signals is due to the fact that they are not saved on disk, and we only had access to their covariances. The results of the extracted $r_0$ from our pipeline against their pipeline is shown in the top part of Figure \ref{fig:regression}. This test shows that the results are not biased due to a software difference. 

\begin{figure}
    \centering
    \includegraphics[width=1\linewidth]{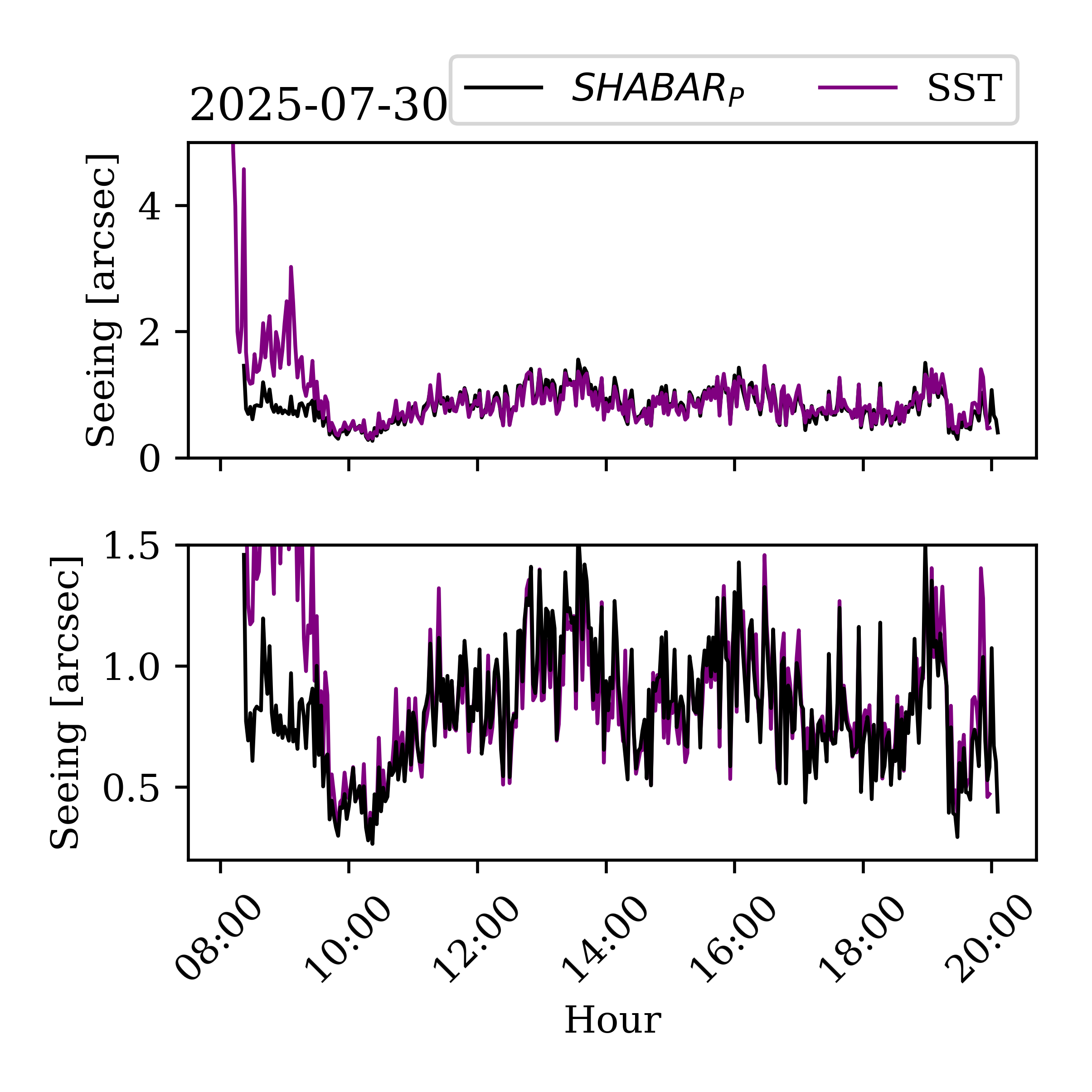}
    \caption{Validation of the $SHABAR_P$ vs $SHABAR_S$. In the upper pannel, the raw seeing results are shown for the $SHABAR_P$ and $SHABAR_S$ (as reduced by SST). In the lower panel, we show a zoom in on the results. Besides the 1$^{st}$ one hour of observation, the two instruments are showing similar results in terms of variation and amplitude. The results are shown for one day of observation (30$^{th}$ of July, 2025).}
    \label{fig:inst_comp}
\end{figure}

The results of comparing both instruments against each other are shown in the bottom pannel of Figure \ref{fig:regression}. It is important to state here that both instruments are working at a different wavelengths that has been corrected when plotting the results ($\lambda_{SST} = 520 nm$, $\lambda_{PoET} = 565 nm$). A direct comparison of the $r_0$, as measured from both instruments, is not straightforward, as the operational wavelength of the two is not equal. As such, to compare them we need to scale the measured $r_0$ to account for that. Using the formula for $r_0$ (Equation 45 in \citet{Hickson2004}), and assuming that the only difference between the two measurements lies in the wavelength, their ratio is given by:

\begin{equation}
    \left(\frac{r0_{PoET}}{r0_{SST}}\right)^{-5/3} = \left(\frac{\lambda_{PoET}}{\lambda_{SST}}\right)^{-2}
    \label{eq:r0}
\end{equation}

Following Equation \ref{eq:r0}, we predict that $r0_{PoET}\sim\ 0.66\ r0_{SST}$. This correction has been applied to the lower pannel of Figure \ref{fig:regression} and to Figure \ref{fig:inst_comp}.

On Figure \ref{fig:inst_comp}, the seeing results obtained from our $SHABAR_P$ are shown along the seeing values from the SST. As it is seen in the upper plot of Figure \ref{fig:inst_comp}, the two instruments are delivering matching results in terms of seeing variation and amplitude, with the exception of the 1$^{st}$ hour of observation. The lower pannel of Figure \ref{fig:inst_comp}, is a zoom in on the results where we show the clear match between the two instruments. Note that the wavelength correction is applied in this plot.


\section{Conclusions}

We have presented the design, implementation, and commissioning of a dedicated daytime seeing monitor to support the operation of the PoET solar telescope. The instrument adopts the SHABAR multi-baseline scintillometer concept and incorporates a fully characterised opto-mechanical and electronic architecture, together with a forward-model inversion pipeline for retrieving \( C_n^2(h) \) profiles and deriving the Fried parameter $r_0$. Laboratory gain measurements validated the stability and uniformity of the detector channels, ensuring consistent normalization across scintillometers. \\

On-sky commissioning at the Swedish Solar Telescope demonstrated that the instrument delivers reliable and temporally coherent estimates of daytime turbulence strength. The retrieved $r_0$ values exhibit the expected diurnal evolution associated with near-ground thermal convection. Comparison with the established SST $SHABAR_S$ system shows that our measurements reproduce both the absolute seeing level and its temporal variability once the wavelength scaling has been applied following Equation \ref{eq:r0}. Pipeline cross-validation using $SHABAR_S$ covariances further confirms that our inversion procedure introduces no systematic bias. \\

These results demonstrate that the $SHABAR_P$ instrument provides robust and operationally suitable daytime seeing diagnostics. Its continuous characterization of the ground-layer turbulence at PoET will enable optimized selection of observing configurations and will contribute to improving the fidelity of high-precision, disk-resolved solar spectroscopy. The results presented in this paper gives us confidence that the $SHABAR_P$ is enough to give us information regarding which aperture size (from 1 to 55 arcseconds) could be used on PoET based on the seeing conditions at the time of observations.

\section*{Acknowledgements}

This work was funded by the European Union (ERC, FIERCE, 101052347). Views and opinions expressed are however those of the authors only and do not necessarily reflect those of the European Union or the European Research Council. Neither the European Union nor the granting authority can be held responsible for them. This work was also supported by FCT - Funda\c{c}$\tilde{a}$o para a Ci\^{e}ncia e a Tecnologia through national funds by these grants: UIDB/04434/2020 DOI: 10.54499/UIDB/04434/2020, UIDP/04434/2020 DOI: 10.54499/UIDP/04434/2020, PTDC/FIS-AST/4862/2020, UID/04434/2025. \\
The authors would like to thank the SST team for all their help and support during the commissioning mission in La Palma.

\section*{Data Availability}

The data underlying this article are based on a commissioning mission between the two instruments. The raw AC and DC files of the $SHABAR_P$ are available upon request from the main author. The data from $SHABAR_S$ is available upon request from the main author. \\
The authors declare no conflict of interest.



\bibliographystyle{rasti}
\bibliography{example} 

@ARTICLE{Wehbe2019,
       author = {{Wehbe}, B. and {Cabral}, A. and {Martins}, J.~H.~C. and {Figueira}, P. and {Santos}, N.~C. and {{\'A}vila}, G.},
        title = "{The impact of atmospheric dispersion in the performance of high-resolution spectrographs}",
      journal = {\mnras},
         year = 2019,
        month = jan,
       volume = {491},
       number = {3},
        pages = {3515-3522},
          doi = {10.1093/mnras/stz3256},
archivePrefix = {arXiv},
       eprint = {1911.08391},
 primaryClass = {astro-ph.IM},
       adsurl = {https://ui.adsabs.harvard.edu/abs/2020MNRAS.491.3515W}
}

@ARTICLE{Wehbe2020a,
       author = {{Wehbe}, B. and {Cabral}, A. and {{\'A}vila}, G.},
        title = "{On-sky measurements of atmospheric dispersion - I. Method validation}",
      journal = {\mnras},
         year = 2020,
        month = nov,
       volume = {499},
       number = {1},
        pages = {183-192},
          doi = {10.1093/mnras/staa2726},
archivePrefix = {arXiv},
       eprint = {2009.01641},
 primaryClass = {astro-ph.IM},
       adsurl = {https://ui.adsabs.harvard.edu/abs/2020MNRAS.499..183W}
}

@ARTICLE{Wehbe2020b,
       author = {{Wehbe}, B. and {Cabral}, A. and {Sbordone}, L. and {{\'A}vila}, G.},
        title = "{On-sky measurements of atmospheric dispersion - II. Atmospheric models characterization}",
      journal = {\mnras},
         year = 2020,
        month = may,
       volume = {503},
       number = {3},
        pages = {3818-3827},
          doi = {10.1093/mnras/stab665},
archivePrefix = {arXiv},
       eprint = {2103.02273},
 primaryClass = {astro-ph.IM},
       adsurl = {https://ui.adsabs.harvard.edu/abs/2021MNRAS.503.3818W}
}

@PROCEEDINGS{Schroeder2000,
        title = "{Astronomical optics}",
    booktitle = {Astronomical optics / Daniel J. Schroeder. San Diego : Academic Press},
         year = 2000,
        editor = {{Schroeder}, Daniel J.},
        month = jan,
       adsurl = {https://ui.adsabs.harvard.edu/abs/2000asop.conf.....S}
}

@ARTICLE{Fried1965,
       author = {{Fried}, D.~L.},
        title = "{Statistics of a Geometric Representation of Wavefront Distortion}",
      journal = {Journal of the Optical Society of America (1917-1983)},
         year = 1965,
        month = nov,
       volume = {55},
       number = {11},
        pages = {1427-1431},
          doi = {10.1364/JOSA.55.001427},
       adsurl = {https://ui.adsabs.harvard.edu/abs/1965JOSA...55.1427F}
}

@BOOK{Hardy1998,
       author = {{Hardy}, John W.},
        title = "{Adaptive Optics for Astronomical Telescopes}",
         year = 1998,
       adsurl = {https://ui.adsabs.harvard.edu/abs/1998aoat.book.....H}
}

@ARTICLE{Santos2000,
       author = {{Santos}, N.~C. and {Mayor}, M. and {Naef}, D. and {Pepe}, F. and {Queloz}, D. and {Udry}, S. and {Blecha}, A.},
        title = "{The CORALIE survey for Southern extra-solar planets. IV. Intrinsic stellar limitations to planet searches with radial-velocity techniques}",
      journal = {\aap},
         year = 2000,
        month = sep,
       volume = {361},
        pages = {265-272},
       adsurl = {https://ui.adsabs.harvard.edu/abs/2000A&A...361..265S}
}

@ARTICLE{Haywood2016,
       author = {{Haywood}, R.~D. and {Collier Cameron}, A. and {Unruh}, Y.~C. and {Lovis}, C. and {Lanza}, A.~F. and {Llama}, J. and {Deleuil}, M. and {Fares}, R. and {Gillon}, M. and {Moutou}, C. and {Pepe}, F. and {Pollacco}, D. and {Queloz}, D. and {S{\'e}gransan}, D.},
        title = "{The Sun as a planet-host star: proxies from SDO images for HARPS radial-velocity variations}",
      journal = {\mnras},
         year = 2016,
        month = apr,
       volume = {457},
       number = {4},
        pages = {3637-3651},
          doi = {10.1093/mnras/stw187},
archivePrefix = {arXiv},
       eprint = {1601.05651},
 primaryClass = {astro-ph.EP},
       adsurl = {https://ui.adsabs.harvard.edu/abs/2016MNRAS.457.3637H}
}

@ARTICLE{Dumusque2015,
       author = {{Dumusque}, Xavier and {Glenday}, Alex and {Phillips}, David F. and {Buchschacher}, Nicolas and {Collier Cameron}, Andrew and {Cecconi}, Massimo and {Charbonneau}, David and {Cosentino}, Rosario and {Ghedina}, Adriano and {Latham}, David W. and {Li}, Chih-Hao and {Lodi}, Marcello and {Lovis}, Christophe and {Molinari}, Emilio and {Pepe}, Francesco and {Udry}, St{\'e}phane and {Sasselov}, Dimitar and {Szentgyorgyi}, Andrew and {Walsworth}, Ronald},
        title = "{HARPS-N Observes the Sun as a Star}",
      journal = {\apjl},
         year = 2015,
        month = dec,
       volume = {814},
       number = {2},
          eid = {L21},
        pages = {L21},
          doi = {10.1088/2041-8205/814/2/L21},
archivePrefix = {arXiv},
       eprint = {1511.02267},
 primaryClass = {astro-ph.EP},
       adsurl = {https://ui.adsabs.harvard.edu/abs/2015ApJ...814L..21D}
}

@ARTICLE{Santos2025,
       author = {{Santos}, N.~C. and {Cabral}, A. and {Leite}, I. and {Smette}, A. and {Abreu}, M. and {Alves}, D. and {Martins}, J.~H.~C. and {Monteiro}, M. and {Silva}, A. and {Wehbe}, B. and {Arancibia}, J. and {{\'A}vila}, G. and {Brillant}, S. and {C{\'a}rdenas}, C. and {Clara}, R. and {Gafeira}, R. and {Gaytan}, D. and {Lovis}, C. and {Miranda}, N. and {Moreno}, P. and {Oliveira}, A. and {Otarola}, A. and {Pepe}, F. and {Rojas}, P. and {Schmutzer}, R. and {Sosnowska}, D. and {van der Heyden}, P. and {Al Moulla}, K. and {Adibekyan}, V. and {Barka}, A. and {Barros}, S.~C.~C. and {Branco}, P. and {Cristo}, E. and {Damasceno}, Y. and {Demangeon}, O. and {Dethier}, W. and {Faria}, J.~P. and {Gomes da Silva}, J. and {Gon{\c{c}}alves}, E. and {Lucero}, J.~P. and {Rodrigues}, J. and {San Nicolas Martinez}, C. and {Santos}, {\^A}. and {Sousa}, S. and {Viana}, P.~T.~P.},
      journal = {The Messenger},
         year = 2025,
        month = mar,
       volume = {194},
        pages = {21-25},
          doi = {10.18727/0722-6691/5381},
       adsurl = {https://ui.adsabs.harvard.edu/abs/2025Msngr.194...21S}
}

@ARTICLE{Griffiths2023,
       author = {{Griffiths}, Ryan and {Osborn}, James and {Farley}, Ollie and {Butterley}, Tim and {Townson}, Matthew J. and {Wilson}, Richard},
        title = "{Demonstrating 24-hour continuous vertical monitoring of atmospheric optical turbulence}",
      journal = {Optics Express},
         year = 2023,
        month = feb,
       volume = {31},
       number = {4},
        pages = {6730},
          doi = {10.1364/OE.479544},
archivePrefix = {arXiv},
       eprint = {2301.07612},
 primaryClass = {astro-ph.IM},
       adsurl = {https://ui.adsabs.harvard.edu/abs/2023OExpr..31.6730G}
}

@ARTICLE{Ozisik2004,
       author = {{{\"O}zi{\v{s}}ik}, T. and {Ak}, T.},
        title = "{First day-time seeing observations at the T{\"U}B{\.I}TAK National Observatory in Turkey}",
      journal = {\aap},
         year = 2004,
        month = aug,
       volume = {422},
        pages = {1129-1133},
          doi = {10.1051/0004-6361:20040326},
       adsurl = {https://ui.adsabs.harvard.edu/abs/2004A&A...422.1129O}
}

@ARTICLE{Seykora1993,
       author = {{Seykora}, E.~J.},
        title = "{Solar Scintillation and the Monitoring of Solar Seeing}",
      journal = {\solphys},
         year = 1993,
        month = jun,
       volume = {145},
       number = {2},
        pages = {389-397},
          doi = {10.1007/BF00690664},
       adsurl = {https://ui.adsabs.harvard.edu/abs/1993SoPh..145..389S}
}

@INPROCEEDINGS{Sliepen2010,
       author = {{Sliepen}, Guus and {J{\"a}gers}, Aswin P.~L. and {Bettonvil}, Felix C.~M. and {Hammerschlag}, Robert H.},
        title = "{Seeing measurements with autonomous, short-baseline shadow band rangers}",
    booktitle = {Ground-based and Airborne Telescopes III},
         year = 2010,
       editor = {{Stepp}, Larry M. and {Gilmozzi}, Roberto and {Hall}, Helen J.},
       series = {Society of Photo-Optical Instrumentation Engineers (SPIE) Conference Series},
       volume = {7733},
        month = jul,
          eid = {77334L},
        pages = {77334L},
          doi = {10.1117/12.857643},
       adsurl = {https://ui.adsabs.harvard.edu/abs/2010SPIE.7733E..4LS}
}

@INPROCEEDINGS{Hill2003,
       author = {{Hill}, F. and {Collados}, M.},
        title = "{Inverting Scintillometer Array Data to Estimate $C_n^{2}(h)$ for the ATST Site Survey}",
    booktitle = {AAS/Solar Physics Division Meeting \#34},
         year = 2003,
       series = {AAS/Solar Physics Division Meeting},
       volume = {34},
        month = may,
          eid = {20.20},
        pages = {20.20},
       adsurl = {https://ui.adsabs.harvard.edu/abs/2003SPD....34.2020H}
}

@INPROCEEDINGS{Wehbe2024,
       author = {{Wehbe}, B. and {Abreu}, M. and {Silva}, A. and {Cabral}, A. and {Santos}, N.},
        title = "{Implementation of a seeing measurement device for the PoET solar telescope}",
    booktitle = {Ground-based and Airborne Instrumentation for Astronomy X},
         year = 2024,
       editor = {{Bryant}, Julia J. and {Motohara}, Kentaro and {Vernet}, Jo{\"e}l. R.~D.},
       series = {Society of Photo-Optical Instrumentation Engineers (SPIE) Conference Series},
       volume = {13096},
        month = jul,
          eid = {1309683},
        pages = {1309683},
          doi = {10.1117/12.3017481},
       adsurl = {https://ui.adsabs.harvard.edu/abs/2024SPIE13096E..83W}
}

@INPROCEEDINGS{Scharmer2003,
       author = {{Scharmer}, Goran B. and {Bjelksjo}, Klas and {Korhonen}, Tapio K. and {Lindberg}, Bo and {Petterson}, Bertil},
        title = "{The 1-meter Swedish solar telescope}",
    booktitle = {Innovative Telescopes and Instrumentation for Solar Astrophysics},
         year = 2003,
       editor = {{Keil}, Stephen L. and {Avakyan}, Sergey V.},
       series = {Society of Photo-Optical Instrumentation Engineers (SPIE) Conference Series},
       volume = {4853},
        month = feb,
        pages = {341-350},
          doi = {10.1117/12.460377},
       adsurl = {https://ui.adsabs.harvard.edu/abs/2003SPIE.4853..341S}
}

@ARTICLE{Pepe2021,
       author = {{Pepe}, F. and {Cristiani}, S. and {Rebolo}, R. and {Santos}, N.~C. and {Dekker}, H. and {Cabral}, A. and {Di Marcantonio}, P. and {Figueira}, P. and {Lo Curto}, G. and {Lovis}, C. and {Mayor}, M. and {M{\'e}gevand}, D. and {Molaro}, P. and {Riva}, M. and {Zapatero Osorio}, M.~R. and {Amate}, M. and {Manescau}, A. and {Pasquini}, L. and {Zerbi}, F.~M. and {Adibekyan}, V. and {Abreu}, M. and {Affolter}, M. and {Alibert}, Y. and {Aliverti}, M. and {Allart}, R. and {Allende Prieto}, C. and {{\'A}lvarez}, D. and {Alves}, D. and {Avila}, G. and {Baldini}, V. and {Bandy}, T. and {Barros}, S.~C.~C. and {Benz}, W. and {Bianco}, A. and {Borsa}, F. and {Bourrier}, V. and {Bouchy}, F. and {Broeg}, C. and {Calderone}, G. and {Cirami}, R. and {Coelho}, J. and {Conconi}, P. and {Coretti}, I. and {Cumani}, C. and {Cupani}, G. and {D'Odorico}, V. and {Damasso}, M. and {Deiries}, S. and {Delabre}, B. and {Demangeon}, O.~D.~S. and {Dumusque}, X. and {Ehrenreich}, D. and {Faria}, J.~P. and {Fragoso}, A. and {Genolet}, L. and {Genoni}, M. and {G{\'e}nova Santos}, R. and {Gonz{\'a}lez Hern{\'a}ndez}, J.~I. and {Hughes}, I. and {Iwert}, O. and {Kerber}, F. and {Knudstrup}, J. and {Landoni}, M. and {Lavie}, B. and {Lillo-Box}, J. and {Lizon}, J.-L. and {Maire}, C. and {Martins}, C.~J.~A.~P. and {Mehner}, A. and {Micela}, G. and {Modigliani}, A. and {Monteiro}, M.~A. and {Monteiro}, M.~J.~P.~F.~G. and {Moschetti}, M. and {Murphy}, M.~T. and {Nunes}, N. and {Oggioni}, L. and {Oliveira}, A. and {Oshagh}, M. and {Pall{\'e}}, E. and {Pariani}, G. and {Poretti}, E. and {Rasilla}, J.~L. and {Rebord{\~a}o}, J. and {Redaelli}, E.~M. and {Santana Tschudi}, S. and {Santin}, P. and {Santos}, P. and {S{\'e}gransan}, D. and {Schmidt}, T.~M. and {Segovia}, A. and {Sosnowska}, D. and {Sozzetti}, A. and {Sousa}, S.~G. and {Span{\`o}}, P. and {Su{\'a}rez Mascare{\~n}o}, A. and {Tabernero}, H. and {Tenegi}, F. and {Udry}, S. and {Zanutta}, A.},
        title = "{ESPRESSO at VLT. On-sky performance and first results}",
      journal = {\aap},
         year = 2021,
        month = jan,
       volume = {645},
          eid = {A96},
        pages = {A96},
          doi = {10.1051/0004-6361/202038306},
archivePrefix = {arXiv},
       eprint = {2010.00316},
 primaryClass = {astro-ph.IM},
       adsurl = {https://ui.adsabs.harvard.edu/abs/2021A&A...645A..96P}
}

@ARTICLE{Klein2024,
       author = {{Klein}, Baptiste and {Aigrain}, Suzanne and {Cretignier}, Michael and {Al Moulla}, Khaled and {Dumusque}, Xavier and {Barrag{\'a}n}, Oscar and {Yu}, Haochuan and {Mortier}, Annelies and {Rescigno}, Federica and {Cameron}, Andrew Collier and {L{\'o}pez-Morales}, Mercedes and {Meunier}, Nad{\`e}ge and {Sozzetti}, Alessandro and {O'Sullivan}, Niamh K.},
        title = "{Investigating stellar activity through eight years of Sun-as-a-star observations}",
      journal = {\mnras},
         year = 2024,
        month = jul,
       volume = {531},
       number = {4},
        pages = {4238-4262},
          doi = {10.1093/mnras/stae1313},
archivePrefix = {arXiv},
       eprint = {2405.12065},
 primaryClass = {astro-ph.EP},
       adsurl = {https://ui.adsabs.harvard.edu/abs/2024MNRAS.531.4238K}
}

@ARTICLE{Zhao2023,
       author = {{Zhao}, Lily L. and {Dumusque}, Xavier and {Ford}, Eric B. and {Llama}, Joe and {Mortier}, Annelies and {Bedell}, Megan and {Al Moulla}, Khaled and {Bender}, Chad F. and {Blake}, Cullen H. and {Brewer}, John M. and {Collier Cameron}, Andrew and {Cosentino}, Rosario and {Figueira}, Pedro and {Fischer}, Debra A. and {Ghedina}, Adriano and {Gonzalez}, Manuel and {Halverson}, Samuel and {Kanodia}, Shubham and {Latham}, David W. and {Lin}, Andrea S.~J. and {Lo Curto}, Gaspare and {Lodi}, Marcello and {Logsdon}, Sarah E. and {Lovis}, Christophe and {Mahadevan}, Suvrath and {Monson}, Andrew and {Ninan}, Joe P. and {Pepe}, Francesco and {Roettenbacher}, Rachael M. and {Roy}, Arpita and {Santos}, Nuno C. and {Schwab}, Christian and {Stef{\'a}nsson}, Gu{\dj}mundur and {Szymkowiak}, Andrew E. and {Terrien}, Ryan C. and {Udry}, Stephane and {Weiss}, Sam A. and {Wildi}, Fran{\c{c}}ois and {Wildi}, Thibault and {Wright}, Jason T.},
        title = "{The Extreme Stellar-signals Project. III. Combining Solar Data from HARPS, HARPS-N, EXPRES, and NEID}",
      journal = {\aj},
         year = 2023,
        month = oct,
       volume = {166},
       number = {4},
          eid = {173},
        pages = {173},
          doi = {10.3847/1538-3881/acf83e},
archivePrefix = {arXiv},
       eprint = {2309.03762},
 primaryClass = {astro-ph.EP},
       adsurl = {https://ui.adsabs.harvard.edu/abs/2023AJ....166..173Z}
}

@ARTICLE{Bouchy2025,
       author = {{Bouchy}, Fran{\c{c}}ois and {Doyon}, Ren{\'e} and {Pepe}, Francesco and {Melo}, Claudio and {Artigau}, {\'E}tienne and {Malo}, Lison and {Wildi}, Fran{\c{c}}ois and {Baron}, Fr{\'e}d{\'e}rique and {Delfosse}, Xavier and {De Medeiros}, Jose Renan and {Rebolo}, Rafael and {Santos}, Nuno C. and {Wade}, Gregg and {Allart}, Romain and {Al Moulla}, Khaled and {Blind}, Nicolas and {Cadieux}, Charles and {Canto Martins}, Bruno L. and {Cook}, Neil J. and {Dumusque}, Xavier and {Frensch}, Yolanda and {Genest}, Fr{\'e}d{\'e}ric and {Gonz{\'a}lez Hern{\'a}ndez}, Jonay I. and {Grieves}, Nolan and {Lo Curto}, Gaspare and {Lovis}, Christophe and {Mignon}, Lucile and {Nielsen}, Louise D. and {Poulin-Girard}, Anne-Sophie and {Rasilla}, Jos{\'e} Luis and {Reshetov}, Vladimir and {Sosnowska}, Danuta and {Sordet}, Michael and {Saint-Antoine}, Jonathan and {Su{\'a}rez Mascare{\~n}o}, Alejandro and {Thibault}, Simon and {Vall{\'e}e}, Philippe and {Vandal}, Thomas and {Abreu}, Manuel and {Aguiar}, Jos{\'e} L.~A. and {Allain}, Guillaume and {Arial}, Tomy and {Auger}, Hugues and {Barros}, Susana C.~C. and {Bazinet}, Luc and {Benneke}, Bj{\"o}rn and {Bonfils}, Xavier and {Boucher}, Anne and {Bourrier}, Vincent and {Bovay}, S{\'e}bastien and {Broeg}, Christopher and {Brousseau}, Denis and {Bruniquel}, Vincent and {Bryan}, Marta and {Cabral}, Alexandre and {Carmona}, Andres and {Carteret}, Yann and {Challita}, Zalpha and {Chazelas}, Bruno and {Cloutier}, Ryan and {Coelho}, Jo{\~a}o and {Cointepas}, Marion and {Conod}, Uriel and {Cowan}, Nicolas B. and {Cristo}, Eduardo and {Gomes da Silva}, Jo{\~a}o and {Dauplaise}, Laurie and {Darveau-Bernier}, Antoine and {de Lima Gomes}, Roseane and {de Freitas}, Daniel Brito and {Delgado-Mena}, Elisa and {Delisle}, Jean-Baptiste and {Ehrenreich}, David and {Faria}, Jo{\~a}o and {Figueira}, Pedro and {Fontinele}, Dasaev O. and {Forveille}, Thierry and {Gagn{\'e}}, Jonathan and {Genolet}, Ludovic and {T{\'e}mich}, F{\'e}lix Gracia and {Hernandez}, Olivier and {Hobson}, Melissa J. and {Hoeijmakers}, Jens and {Hubin}, Norbert and {Jahandar}, Farbod and {Jayawardhana}, Ray and {K{\"a}ufl}, Hans-Ulrich and {Kerley}, Dan and {Kolb}, Johann and {Krishnamurthy}, Vigneshwaran and {Lafreni{\`e}re}, David and {Lamontagne}, Pierrot and {Larue}, Pierre and {Leath}, Henry and {L'Heureux}, Alexandrine and {de Castro Le{\~a}o}, Izan and {Lim}, Olivia and {Martins}, Allan M. and {Matthews}, Jaymie and {Mayer}, Jean-S{\'e}bastien and {Messias}, Yuri S. and {Metchev}, Stan and {Moranta}, Leslie and {Mordasini}, Christoph and {Mounzer}, Dany and {Nari}, Nicola and {Osborn}, Ares and {Ouellet}, Mathieu and {Otegi}, Jon and {Parc}, L{\'e}na and {Pasquini}, Luca and {Passegger}, Vera M. and {Pelletier}, Stefan and {Peroux}, C{\'e}line and {Piaulet-Ghorayeb}, Caroline and {Plotnykov}, Mykhaylo and {Pompei}, Emanuela and {Rowe}, Jason and {Sarajlic}, Mirsad and {Segovia}, Alex and {Seidel}, Julia and {S{\'e}gransan}, Damien and {Schnell}, Robin and {Costa Silva}, Ana Rita and {Srivastava}, Avidaan and {Stefanov}, Atanas K. and {Teixeira}, M{\'a}rcio A. and {Udry}, St{\'e}phane and {Valencia}, Diana and {Vaulato}, Valentina and {Wardenier}, Joost P. and {Wehbe}, Bachar and {Weisserman}, Drew and {Wevers}, Ivan and {Yariv}, Vincent and {Zins}, G{\'e}rard},
        title = "{NIRPS joining HARPS at ESO 3.6 m: On-sky performance and science objectives}",
      journal = {\aap},
         year = 2025,
        month = aug,
       volume = {700},
          eid = {A10},
        pages = {A10},
          doi = {10.1051/0004-6361/202453341},
archivePrefix = {arXiv},
       eprint = {2507.21767},
 primaryClass = {astro-ph.IM},
       adsurl = {https://ui.adsabs.harvard.edu/abs/2025A&A...700A..10B}
}

@ARTICLE{Hall2018,
       author = {{Hall}, Richard D. and {Thompson}, Samantha J. and {Handley}, Will and {Queloz}, Didier},
        title = "{On the Feasibility of Intense Radial Velocity Surveys for Earth-Twin Discoveries}",
      journal = {\mnras},
         year = 2018,
        month = sep,
       volume = {479},
       number = {3},
        pages = {2968-2987},
          doi = {10.1093/mnras/sty1464},
archivePrefix = {arXiv},
       eprint = {1806.00518},
 primaryClass = {astro-ph.EP},
       adsurl = {https://ui.adsabs.harvard.edu/abs/2018MNRAS.479.2968H}
}

@INPROCEEDINGS{Thompson2016,
       author = {{Thompson}, Samantha J. and {Queloz}, Didier and {Baraffe}, Isabelle and {Brake}, Martyn and {Dolgopolov}, Andrey and {Fisher}, Martin and {Fleury}, Michel and {Geelhoed}, Joost and {Hall}, Richard and {Gonz{\'a}lez Hern{\'a}ndez}, Jonay I. and {ter Horst}, Rik and {Kragt}, Jan and {Navarro}, Ram{\'o}n and {Naylor}, Tim and {Pepe}, Francesco and {Piskunov}, Nikolai and {Rebolo}, Rafael and {Sander}, Louis and {S{\'e}gransan}, Damien and {Seneta}, Eugene and {Sing}, David and {Snellen}, Ignas and {Snik}, Frans and {Spronck}, Julien and {Stempels}, Eric and {Sun}, Xiaowei and {Santana Tschudi}, Samuel and {Young}, John},
        title = "{HARPS3 for a roboticized Isaac Newton Telescope}",
    booktitle = {Ground-based and Airborne Instrumentation for Astronomy VI},
         year = 2016,
       editor = {{Evans}, Christopher J. and {Simard}, Luc and {Takami}, Hideki},
       series = {Society of Photo-Optical Instrumentation Engineers (SPIE) Conference Series},
       volume = {9908},
        month = aug,
          eid = {99086F},
        pages = {99086F},
          doi = {10.1117/12.2232111},
archivePrefix = {arXiv},
       eprint = {1608.04611},
 primaryClass = {astro-ph.IM},
       adsurl = {https://ui.adsabs.harvard.edu/abs/2016SPIE.9908E..6FT}
}

@ARTICLE{Davies2012,
       author = {{Davies}, Richard and {Kasper}, Markus},
        title = "{Adaptive Optics for Astronomy}",
      journal = {\araa},
         year = 2012,
        month = sep,
       volume = {50},
        pages = {305-351},
          doi = {10.1146/annurev-astro-081811-125447},
archivePrefix = {arXiv},
       eprint = {1201.5741},
 primaryClass = {astro-ph.IM},
       adsurl = {https://ui.adsabs.harvard.edu/abs/2012ARA&A..50..305D}
}

@article{Rao2024,
  author       = {Rao, Changhui and Zhong, Libo and Guo, Youming and Li, Min and Zhang, Lanqiang and Wei, Kai},
  title        = {Astronomical adaptive optics: a review},
  journal      = {PhotoniX},
  volume       = {5},
  pages        = {16},
  year         = {2024},
  doi          = {10.1186/s43074-024-00118-7},
  url          = {https://photonix.springeropen.com/articles/10.1186/s43074-024-00118-7}
}

@INPROCEEDINGS{Beckers2003,
       author = {{Beckers}, Jacques M. and {Liu}, Zhong and {Jin}, Zhenyu},
        title = "{ATST seeing monitor: February 2002 observations at Fuxian Lake}",
    booktitle = {Innovative Telescopes and Instrumentation for Solar Astrophysics},
         year = 2003,
       editor = {{Keil}, Stephen L. and {Avakyan}, Sergey V.},
       series = {Society of Photo-Optical Instrumentation Engineers (SPIE) Conference Series},
       volume = {4853},
        month = feb,
        pages = {273-284},
          doi = {10.1117/12.460268},
       adsurl = {https://ui.adsabs.harvard.edu/abs/2003SPIE.4853..273B}
}

@ARTICLE{Beckers1993,
       author = {{Beckers}, Jacques M.},
        title = "{On the Relation Between Scintillation and Seeing Observations of Extended Objects}",
      journal = {\solphys},
         year = 1993,
        month = jun,
       volume = {145},
       number = {2},
        pages = {399-402},
          doi = {10.1007/BF00690665},
       adsurl = {https://ui.adsabs.harvard.edu/abs/1993SoPh..145..399B}
}

@ARTICLE{Beckers1997,
       author = {{Beckers}, Jacques M. and {Leon}, Ed and {Mason}, Jim and {Wilkins}, Larry},
        title = "{Solar Scintillometry: Calibration of Signals and its Use for Seeing Measurements}",
      journal = {\solphys},
         year = 1997,
        month = nov,
       volume = {176},
       number = {1},
        pages = {23-36},
          doi = {10.1023/A:1004988028580},
       adsurl = {https://ui.adsabs.harvard.edu/abs/1997SoPh..176...23B}
}

@ARTICLE{Hickson2004,
       author = {{Hickson}, Paul and {Lanzetta}, Kenneth},
        title = "{Measuring Atmospheric Turbulence with Lunar Scintillometer Array}",
      journal = {\pasp},
         year = 2004,
        month = dec,
       volume = {116},
       number = {826},
        pages = {1143-1152},
          doi = {10.1086/427046},
       adsurl = {https://ui.adsabs.harvard.edu/abs/2004PASP..116.1143H}
}




\appendix

\section{Scintillometer specifications}
\label{app:scint}
The components of the scintillometer unit were chosen off-the-shelf to minimize the cost of the instrument. However, the PCB was custom made taken into consideration the amount of light expected and the chosed photodiodes. The housing of each unit was custom made to host the PCB and the optical tube containing the lenses, filtes, and photodiodes. The separation between the scintillometers (Table \ref{tab:positions}) follow the same proposed distances as in \citet{Sliepen2010}. The off-the-shelf components are listed in Table \ref{tab:scintillometer}. The full detailed design can be accessed via request from the authors.
\begin{table}
    \centering
    \begin{tabular}{c|c}
       Scintillometer unit  & Position (mm)  \\
       \hline
        0 & 0 \\
        1 & 20.0 \\
        2 & 55.2 \\
        3 & 117.7 \\
        4 & 228.1 \\
        5 & 423.4 \\
        \hline
    \end{tabular}
    \caption{Positions of the scintillometers relative to the first scintillometer at the start of the bar. The separations are similar to the ones used in the instrument described in \citet{Sliepen2010}.}
    \label{tab:positions}
\end{table}

\begin{table}
    \centering
    \begin{tabular}{c|c|c}
        Component & Company & Reference \\
        \hline
        ND Filter & Thorlabs & NDUV509B \\
        Lens & Thorlabs & LA1560 \\
        BP Filter & Edmund Optics & 18-501 \\
        Field Stop & Custom made & -- \\
        Photodiode & Vishay & BPW21R \\
        Optical tube & Thorlabs & SM05L05-SM05L03 \\
        \hline
    \end{tabular}
    \caption{Details of the scintillometer components}
    \label{tab:scintillometer}
\end{table}

The results shown in Table \ref{tab:gains} and mentioned in subsection \ref{subsec:labtests} validate the stability and uniformity of the scintillometrs channels.

\begin{table}
    \centering
    \begin{tabular}{c|c|c|c|c}
       Unit  & $DC_{gain}$ & $AC_{gain\times 10}$ & $AC_{gain\times 50}$ & $AC_{gain\times 100}$\\
       \hline
        Box 0	&	14720	&	130769	&	615385	&	1238462 \\
        Box 1	&	14950	&	134615	&	623077	&	1230769 \\
        Box 2	&	13470	&	130769	&	630769	&	1246154 \\
        Box 3	&	14880	&	136154	&	623077	&	1261538 \\
        Box 4	&	14920	&	138462	&	630769	&	1261538 \\
        Box 5	&	14820	&	134615	&	615385	&	1269231 \\
    
        \hline
 
    \end{tabular}
    \caption{Results of the gain quantification of each scintillometer unit. We show the results for the DC gain, as well as the different AC gains based on the jumper position.}
    \label{tab:gains}
\end{table}



\bsp	
\label{lastpage}
\end{document}